\documentclass[12pt,a4paper,oneside,reqno]{article}

\usepackage{amsmath,amssymb}
\usepackage{color}
\usepackage{endnotes}

\newcommand{\ax}[1]{\textcolor{axcolor}{\ensuremath{\mathsf{#1}}}}

\definecolor{axcolor}{rgb}{.4,0,.4}
\definecolor{axbgcolor}{rgb}{1,.7,1}

\newcommand{\Ph}{\mathsf{Ph}}
\newcommand{\Q}{\mathit{Q}}
\newcommand{\B}{\mathit{B}}
\newcommand{\W}{\mathsf{W}}
\newcommand{\IOb}{\mathsf{IOb}}

\begin{document}

\makeatletter
 \renewcommand\section{\@startsection {section}{1}{\z@}%
                     {-3.25ex\@plus -1ex \@minus -.2ex}%
                     {1.5ex \@plus .2ex}%
                     {\normalfont\scshape\centering}} 
\makeatother

\title{On Why-Questions in Physics}
\author{Gergely Sz\'ekely}
\date{}

\maketitle

\section{introduction}
\footnotetext{This research is supported by the Hungarian National
  Foundation for Scientific Research grant No T73601.  } In natural
sciences, the most interesting and relevant questions are the
so-called why-questions. What is a why-question?  A why-question is
nothing else than a question in the form ``Why $P$?'' (or ``Why is $P$
true?'') where $P$ is an arbitrary statement.

There are several different approaches to why-questions and
explanations in the literature, see, e.g., \cite{friedman},
\cite{hempel-oppenheim}, \cite{hintikka-haloen}, \cite{temple}.
However, most of the literature deals with why-questions about
particular events, such as ``Why did Adam eat the apple?''. Even the
best known theory of explanation, Hempel's covering law model, is
designed for explaining particular events.  Here we only deal with
purely theoretical why-questions about general phenomena of physics,
for instance ``Why can no observer move faster than light?'' or ``Why
are Kepler's laws valid?''.

Here we are not going to develop a whole new theory of why-questions
in physics. We will just touch upon some ideas and examples relevant
to our subject.

\section{answering why-questions}

How to answer a why-question?  For example, let NoFTL be the statement
``No observer can move faster than light,'' which is one of the
several astonishing predictions of relativity. As it is a statement
that hard to believe, it is natural to ask why we think it is true.
The standard answers to the question ``Why NoFTL?'' are
\begin{enumerate}
\item ``NoFTL is true because the $4$-dimensional Minkowski spacetime
  over $\mathbb{R}$ (the field of real numbers) is a good model of our
  physical world; and in this model NoFTL is valid.''
\item ``NoFTL is an axiom (of Special Relativity).''
\end{enumerate}

Neither of these answers is satisfactory to a logician. The problem
with the first is that it refers to one particular model and not to a
list of axioms.  The second does not really answer the question, it is
a kind of ``just because'' answer.

How to give satisfactory answers to why-questions in physics?  First
of all to answer the question ``Why $P$?'', we need a formal language
in which we can formulate $P$ and the possible answers to it. Let us
fix one such language.\footnote{Every concept we introduce here is
  relative to a fixed language in which $P$ can be formulated.}  The
{\it possible answers} to the question ``Why $P$?'' are consistent
theories which do not contain $P$ as an axiom.\footnote{It is important
  that we restrict our definition to possible answers to why-questions
  in physics because, for example, in mathematics the possible answers to
  why-questions are usually proofs and not theories. See, e.g.,
  \cite{sandborg}.} Here we do not require theories to be deductively
closed systems, as we would like to separate the assumptions of the
theory from its consequences. Hence we use the term theory as a
synonym for axiom system. Let us call a possible answer {\it
  acceptable answer} if it implies $P$.

To present some examples of acceptable answers to the question ``Why NoFTL?'', let us
consider the following two-sorted first-order language:
\[\{\,\Q,+,\cdot,<; \B, \IOb, \Ph; \W\,\},\]
 where $\Q$ is the sort of quantities and $\B$ is the sort of bodies;
 $\cdot$ and $+$ are binary function symbols and $<$ is a binary
 relation symbol of sort $\Q$; $\IOb$ (inertial observers) and
 $\Ph$ (light signals or photons) are unary relation symbols of
 sort $\B$; and $\W$ (the world-view relation) is a $6$-ary
 relation symbol of the type
 $\B\times\B\times\Q\times\Q\times\Q\times\Q$.  Relations $\IOb(o)$
 and $\Ph(b)$ are translated as ``$o$ is an inertial observer,'' and
 ``$b$ is a photon,'' respectively.  We use the world-view relation
 $\W$ to speak about coordinatization by translating $\W(o,b,x,y,z,t)$
 as ``observer $o$ coordinatizes body $b$ at spacetime location
 $\langle x,y,z,t\rangle$,'' (that is, at space location $\langle
 x,y,z\rangle$ and at instant $t$).

In this language we can define the term $speed_o(b)$ as the speed of a
body $b$ according to observer $o$. Hence we can formulate that no
observer can move faster than light. Moreover, we can prove the following:
\[
\ax{SpecRel}\models \forall
o,o',p\enskip \IOb(o)\land\IOb(o')\land\Ph(p)\implies
speed_{o}(o')<speed_{o}(p),
\]
where \ax{SpecRel} is the consistent axiom system of the following axioms:
\begin{description}
\item[\ax{AxField}:] The {\it quantity part} $\langle \Q;+,\cdot, <
  \rangle$ is an ordered field.
\item[\ax{AxSelf}:] Every observer coordinatizes itself at a coordinate point
  if and only if its space component is the origin, that is, space location $\langle 0,0,0\rangle$.
\item[\ax{AxPh}:] The speed of light signals is $1$ according to every
  inertial observer.
\item[\ax{AxEv}:] Every {inertial} observer coordinatizes the
  very same set of events.
\item[\ax{AxSymd}:] {Inertial} observers agree as for the spatial
  distance between events if these events are simultaneous for both of them.
\end{description}
For the formulation of these concepts and axioms, see, e.g., \cite{wku-amnsz}.
So \ax{SpecRel} is an acceptable answer to the question
``Why NoFTL?'', see \cite[Thm.1]{wku-amnsz}.

\section{how to compare the different answers}

How to get better and better answers to a certain why-question? The
basic idea is that ``the less it assumes, the better an answer is.''
Let us try to make this idea more precise by introducing the following concepts:
\begin{enumerate}
\item $Th_2$ is {\it nonworse} than $Th_1$ as an answer  to the question ``Why $P$?''
   if all the formulas of $Th_2$ are consequences of
  $Th_1$, that is, $Th_1$ implies $\varphi_2$ for any formula
  $\varphi_2\in Th_2$.
\item $Th_2$ is {\it piecewise nonworse} than $Th_1$ as an answer to the question ``Why
  $P$?''  if for any formula $\varphi_2\in Th_2$ there is
  $\varphi_1\in Th_1$ such that $\varphi_1$ implies $\varphi_2$.
\end{enumerate}
It is easy to see the following: 
\begin{enumerate}
\item Both  concepts above are preorders on sets of
  formulas, that is, they are transitive and reflexive relations.
\item $Th_2$ is a piecewise nonworse answer than $Th_1$ if
  $Th_2\subseteq Th_1$.
\item $Th_2$ is a nonworse answer than $Th_1$ if it is piecewise
  nonworse.
\end{enumerate}
Let us say that $Th_2$ is {\it (piecewise) better} than $Th_1$ if it is
(piecewise) nonworse and nonequivalent according to the equivalence
relation defined by the corresponding preorder.

These definitions fit in with the following idea of Michael Friedman \cite{friedman}:
 ``Science
increases our understanding of the world by reducing the total number
of independent phenomena that we have to accept as ultimate or
given.''

To present another example, let us note that the following theorem can also be proved:  
\[
\ax{SpecRel_0}\models \forall 
o,o',p\enskip \IOb(o)\land\IOb(o')\land\Ph(p)\implies
speed_{o}(o')<speed_{o}(p),
\]
where \ax{SpecRel_0} consists of the following four axioms only:
\ax{AxField}, \ax{AxSelf}, \ax{AxPh} and \ax{AxEv}.  Hence
\ax{SpecRel_0} is a piecewise better answer to the question ``Why
NoFTL?'' than \ax{SpecRel}. Further answers to this question, which
are not comparable to \ax{SpecRel_0} according to the preorders
above, can be found in \cite[Thms.\ 3 and 5]{MNT}.

Let us note that, according to the definition above, a theory
consisting of two (nonequivalent) axioms is a piecewise better answer
than a theory consisting of their conjunction. That is a nice
property as it makes the introduction of the following concept possible.
Let us say that a possible answer to the question ``Why $P$?'' is {\it
  pointless} if there is a piecewise better answer which contains $P$
as an axiom. Accordingly, the conjunction of Kepler's
laws and Boyle's law is a pointless answer to the question ``Why
Kepler's laws are valid?''.  So it eliminates a problem that motivated
Hempel to introduce his covering law model only for the explanation of
particular events, see \cite{hempel-oppenheim}.

\section{general answers}

In the previous section, we dealt with answers to one particular
question. However, in physics it is a general desire to search for
unified theories, that is, theories answering more questions. So a
good answer in physics is characterized by assuming little, but
implying a lot.\footnote{Of course there are other
  desired requirements of a physical theory, such as simplicity and
  comprehensibility of the axioms or experimental testability.  However, it
  is not easy to  define these concepts precisely if it is
  possible at all.}

The unification of theories is the point where we have to leave the
convenience of fixed languages and search for suitable unification of
the languages of the theories in question, too.  For example, the
juxtaposition of the axioms of relativity and quantum theories in
their combined language will result in one theory but it will not
solve the problem of their reconciliation. Though it will be
consistent (as its parts are consistent) and will imply all the
prediction of relativity and quantum theories, it will not be what we
mean by a unified theory. To achieve a truly unified theory, we need a
richer language in which we can formulate the interrelations between
the concepts of these theories. However, unifying the languages in an
appropriate way and formulating new axioms in this unified language to
establish the interrelations between their concepts such that the
unified theory has new experimentally testable predictions is a very
difficult task.  Unifying theories is the subject of amalgamating
theories in algebraic logic, see, e.g., \cite{madarasz}.

\section{the truth of axioms of physics}

Most theories of why-questions require the explanations to be true,
see, e.g., \cite{hempel-oppenheim}. However, it is a
fundamental requirement of a physical theory to be experimentally
testable, hence refutable. So in physics we do not know whether an axiom is true or not, we
just presume so.  Therefore, if we require the explanations to be
true, we will never ascertain whether our theory is really an explanation or
not. Hence in the case of why-questions of physics, it is better not
to require the axioms to be true.

For example, if we required the axioms of physics to be true, we could not
say that Newton's theory is an explanation of Kepler's laws because Newton's
theory is refuted by some experimental facts.\footnote{For example, the
  perihelion advance of Mercury is a well known experimental fact that
  disproves Newton's theory.} That would be inconvenient as Newton's theory
is the standard example in the literature for explanation of
Kepler's laws. Hence it is better to treat the axioms as possible truths
and treat the matter only in conditional, that is, an
answer to the question ``Why $P$?'' means that ``If the theory is true,
then so is $P$.'' In this sense it is still meaningful to say that
Newton's theory explains Kepler's laws. And according to the above definition Newton's
theory is an acceptable answer to the question ``Why Kepler's
laws are valid?''.

\section{some further examples}
We can take the Twin Paradox (TwP) and ask ``Why TwP?''.  In the
language of \ax{SpecRel}, we cannot formulate the full version of TwP
only its inertial approximation called the Clock Paradox (ClP). Here
we only concentrate on TwP  but for similar investigation of ClP see
\cite{myphd}, \cite{gcocp}.  To formulate the full version of TwP we
have to extend the language above for accelerated (that is,
non-inertial) observers, which is done by adding one more unary relation
for accelerated observers on the sort of bodies. In this language we
can formulate TwP, see \cite{twp}, \cite{myphd}. Let us
denote the formulated version of TwP as \ax{TwP}.

The most natural axiom to assume about accelerated observers is the following:
\begin{description}
\item[\ax{AxCmv}:] At each moment of its world-line, every accelerated observer
  coordinatizes the nearby world for a short while in the same way as an {inertial
    observer} does.
\end{description}
For precise formulation of this axiom, see  \cite{twp}, \cite{myphd}.
Let \ax{AccRel} be the axiom system consisting of \ax{AxCmv} and all the five
axioms of \ax{SpecRel}.

Surprisingly, \ax{AccRel} does not answer the question ``Why
TwP?''. Moreover, the following can be proved:
\[
\ax{AccRel}\cup\ax{Th(\mathbb{R})}\not\models \ax{TwP},
\]
where \ax{Th(\mathbb{R})} is the whole first-order theory of real
numbers, see \cite{twp}.  At first sight this result suggests that the
question ``Why TwP?'' cannot be answered within first-order logic,
which would be depressing as there are weighty methodological reasons
for staying within first-order logic, see, e.g., \cite{pezsgo},
\cite{myphd}.  However, there is a first-order axiom scheme (\ax{IND})
in the above language such that \ax{AccRel} together with this scheme
answers the question ``Why TwP?'', see \cite{twp}, \cite{myphd}.

The why-question ``Why gravity slows time down?'', can be answered by the
theory $\ax{AccRel}\cup\ax{IND}$ and Einstein's equivalence principle,
see \cite{MNSz}, \cite{myphd}.

For further examples, let us consider the question of ``Why does
relativistic mass increase?'' or ``Why the sum of the rest masses of
inelastically colliding inertial bodies is smaller than the rest mass of
the originated inertial body?''.  Theories \ax{SpecRelDyn} and
\ax{SpecRelDyn^+} are possible answers to these questions,
respectively, see \cite{dyn-studia}, \cite{wku-msz}, \cite{myphd}.

\section{concluding remarks}

In the spirit of reverse mathematics, we have introduced a precise
definition of acceptable answers to why-questions in physics and some
ideas about how to compare these answers. We also presented several examples
mainly from axiomatic relativity. Finally let us note that the work
done by the research group of Logic and Relativity led by
Andr\'eka Hajnal and Istv\'an N\'emeti can be considered as providing
explanations to why-questions of relativity, see references.


Alfr\'ed R\'enyi Institute of Mathematics\\
of the Hungarian Academy of Sciences\\
POB 127, H-1364 Budapest, Hungary\\
turms@renyi.hu


\begin{thebibliography}{99}
\footnotesize
\bibitem{AMNsamples} Hajnal Andr{\'e}ka, Judit X.~Madar{\'a}sz, and
  Istv{\'a}n N{\'e}meti.  ``Logical axiomatizations of
  space-time. Samples from the literature'', in: {\it Non-Euclidean
    geometries}, volume 581 of {\it Mathematics and Its Applications}. New York:
  Springer, 2006, pp.155-185. 

\bibitem{logst} Hajnal Andr{\'e}ka, Judit X.~Madar{\'a}sz, and
  Istv{\'a}n N{\'e}meti, ``Logic of space-time and relativity theory'',
  in: Marco Aiello, Ian Pratt-Hartmann and Johan van Benthem (Eds.),
  {\it Handbook of Spatial Logics}. Dordrecht: Springer, 2007,
  pp.607-711.

\bibitem{pezsgo} Hajnal Andr{\'e}ka, Judit X.~Madar{\'a}sz, and
  Istv{\'a}n N{\'e}meti,  {\it On the logical structure of relativity
    theories}.  Research report, Alfr{\'e}d R{\'e}nyi Institute of
  Mathematics, Budapest, 2002.  With
  contributions from Attila Andai, G{\'a}bor S{\'a}gi, Ildik{\'o} Sain
  and Csaba
  T{\H o}ke. http://www.math-inst.hu/pub/algebraic-logic/Contents.html. 1312
  pp. 

\bibitem{dyn-studia} Hajnal Andr{\'e}ka, Judit X.~Madar{\'a}sz,
  Istv{\'a}n N{\'e}meti, and Gergely Sz{\'e}kely, ``Axiomatizing
  relativistic dynamics without conservation postulates'', in: {\it Studia
    Logica} 89, 2, 2008, pp.163-186.

\bibitem{wku-amnsz} Hajnal Andr{\'e}ka, Judit X.~Madar{\'a}sz,
  Istv{\'a}n N{\'e}meti, and Gergely Sz{\'e}kely, Vienna Circle and
  Logical Analysis of Relativity Theory, in: this volume.


\bibitem{friedman} Michael Friedman, ``Explanation and Scientific
  Understanding'', in: {\it The Journal of Philosophy} 71, 1, 1974,
  pp.5-19.

\bibitem{hempel-oppenheim} Carl G. Hempel and Paul Oppenheim,
  ``Studies in the Logic of Explanation'', in: {\it Philosophy of
  Science} 15, 2, 1948, pp.135-175.

\bibitem{hintikka-haloen} Jaakko Hintikka and Ilpo Halonen,
  ``Semantics and Pragmatics for Why-Questions'', in: {\it The Journal
  of Philosophy} 92, 12, 1995, pp.636-657.

      
\bibitem{Mphd} Judit X.~Madar{\'a}sz.  {\em Logic and Relativity (in
  the light of definability theory)}.  PhD thesis, E{\"o}tv{\"o}s
  Lor{\'a}nd Univ., Budapest, 2002. 

\bibitem{madarasz} Judit X.~Madar{\'a}sz. ``Amalgamation and
Interpolation; Pushing the limits.'' Parts I-II, in: {\it Studia Logica} 61, 3, 1998,
pp.311-345 and 62, 1, 1999, pp.1-19.


\bibitem{twp} Judit X.~Madar{\'a}sz, Istv{\'a}n N{\'e}meti, and Gergely
  Sz{\'e}kely, ``Twin paradox and the logical foundation of relativity
  theory'', in: {\it Foundations of Physics} 36, 5, 2006, pp.681-714.

\bibitem{MNSz} Judit X.~Madar{\'a}sz, Istv{\'a}n N{\'e}meti, and Gergely
  Sz{\'e}kely, ``First-order logic foundation of relativity theories'',
  Dov Gabbay, Sergei Goncharov and Michael Zakharyaschev (Eds.), in: {\it
    Mathematical problems from applied logic II}. New York: Springer,
  2007, pp.217-252.


\bibitem{MNT} Judit X.~Madar{\'a}sz, Istv\'an N{\'e}meti, and Csaba
  T{\H o}ke, ``On generalizing the logic-approach to space-time
  towards general relativity: first steps'', in: Vincent F.\ Hendricks,
  Fabian Neuhaus, Stig Andur Pedersen, Uwe Scheffler and Heinrich Wansing
  (Eds.), {\it First-Order Logic Revisited}.  Berlin: Logos Verlag
  2004, pp.225-268.

\bibitem{wku-msz} Judit X.~Madar{\'a}sz,
   and Gergely Sz{\'e}kely, Comparing Relativistic and Newtonian
    Dynamics in First-Order Logic, in: this volume.


\bibitem{sandborg} David Sandborg, ``Mathematical Explanation and the
  Theory of Why-Questions'', in: {\it The British Journal for the
    Philosophy of Science}, 49, 4, 1998, pp.603-624.
  
\bibitem{myphd} Gergely Sz{\'e}kely, {\it First-Order Logic
  Investigation of Relativity Theory with an Emphasis on Accelerated
  Observers}. PhD thesis,  E{\"o}tv{\"o}s
  Lor{\'a}nd Univ., Budapest, 2009.


\bibitem{gcocp} Gergely Sz{\'e}kely, ``Geometrical characterization of
  the twin paradox and its variants'', submitted. 

\bibitem{temple} Dennis Temple, ``The Contrast Theory of
  Why-Questions'', in: {\it Philosophy of Science} 55, 1, 1988,
  pp.141-151.

\end{thebibliography}
\end{document}